\documentclass[reprint,final,nofootinbib]{revtex4-1}
\pdfoutput=1
\usepackage{amssymb} 
\usepackage{amsthm}
\usepackage{amsmath}
\usepackage{graphicx}
\usepackage{subcaption}
\usepackage{dcolumn}
\usepackage{bm}
\usepackage{cancel}
\usepackage{float}
\usepackage{epstopdf}
\usepackage{footnote}
\usepackage{url}
\usepackage[svgnames] {xcolor}
\usepackage{hyperref}

\begin{document}
\title{Lunar antineutrinos and heat: fluxes from primordial radioactivity}

\author{S. T. Dye and A. M. Barna}
\affiliation{Ronin Institute, Montclair, NJ, USA}
\date{\today}
\begin{abstract}
\vspace{1mm}
\noindent
We estimate the fluxes of heat and antineutrinos due to primordial radioactivity within the moon. We use a radial density profile, specifying an inner core and a model-averaged crust. Thickness, density, and elevation of the lunar crust are from remote measurements of the gravitational field. Lateral and vertical variations of thorium, uranium, and potassium abundances in the crust follow from a prediction of the lunar bulk chemical composition. We constrain the total contents of thorium, uranium, and potassium using estimates for the earth's primitive mantle. These contents produce $311\pm37$ GW of radiogenic heating and a surface-averaged heat flux of $8.19\pm0.97$ mW/m$^2$. Our lunar model estimates an antineutrino flux of $(1.83\pm0.32)\times10^6$  cm$^{-2}$s$^{-1}$ and an antineutrino inverse beta decay rate of $5.8\pm1.0$ per $10^{32}$ free proton targets per year, both averaged over the surface.
\end{abstract}

\maketitle
\section{Introduction}
Long-lived nuclides of thorium, uranium, and potassium ($^{232}$Th, $^{238}$U, $^{235}$U, $^{40}$K) in the silicate layers of the moon produce fluxes of heat and antineutrinos. These fluxes depend on the thicknesses, densities, and radiogenic contents of the lunar crust, mantle, and low viscosity zone (LVZ). We build a model of the moon using the characteristics of a radial density profile \cite{briaud_etal2023}, gridded data on crustal thickness, density, and elevation derived from GRAIL gravity measurements \cite{wieczorek_etal2013}, and a prediction of lateral and vertical abundances of thorium in the crust \cite{taylor_wieczorek_2014}. Our lunar model estimates the fluxes of heat and antineutrinos, which vary across the lunar surface. We compare our estimates of the heat flux with measurements at three sites and find good agreement. The predictions from our model are potentially useful for selecting the location of future lunar heat flux measurements or sample collection sites \cite{siegler_etal2022}. Our estimates of the antineutrino fluxes allow calculation of interaction rates from various targets, including free protons, atomic electrons, and selected atomic nuclei. We include the antineutrino flux estimates from our lunar model in a web application, which calculates the various antineutrino interaction rates at any location on the moon's surface. Interaction rates inform potential antineutrino detection projects with applications for testing models of the moon's origin and thermal evolution, as well as for extraterrestrial nuclear monitoring and security  \cite{bernstein_etal2020}.

\section{Physical Model}
\begin{figure}
\centering
\includegraphics[trim = 0mm 0mm 0mm 0mm, clip, scale=0.40]{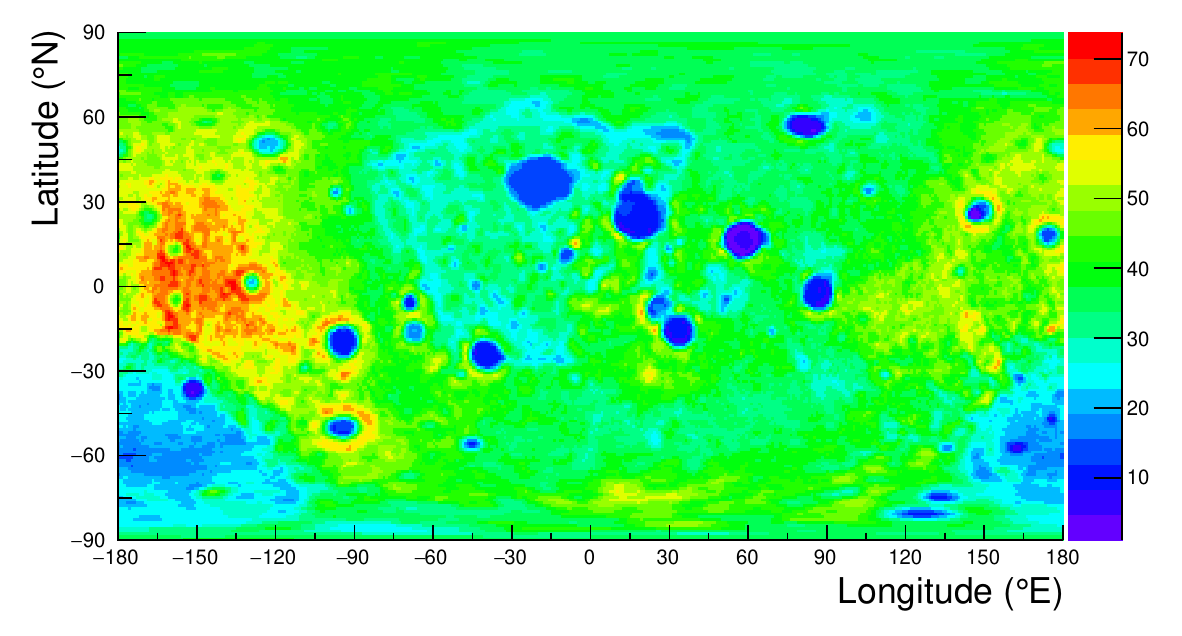}
\caption{Map of the lunar crust thickness (km).}
\label{fig:lunar_crust_thickness}
\end{figure}

\begin{figure}
\centering
\includegraphics[trim = 0mm 0mm 0mm 0mm, clip, scale=0.40]{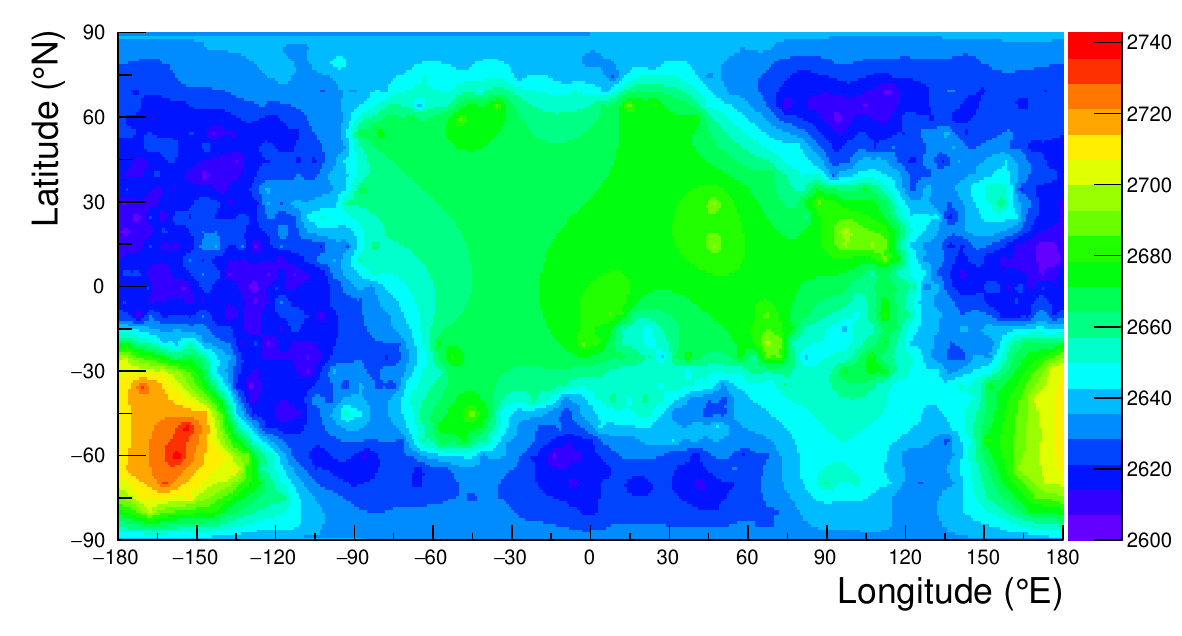}
\caption{Map of the lunar crust density (kg/m$^3$).}
\label{fig:lunar_crust_density}
\end{figure}

\begin{figure}
\centering
\includegraphics[trim = 0mm 0mm 0mm 0mm, clip, scale=0.40]{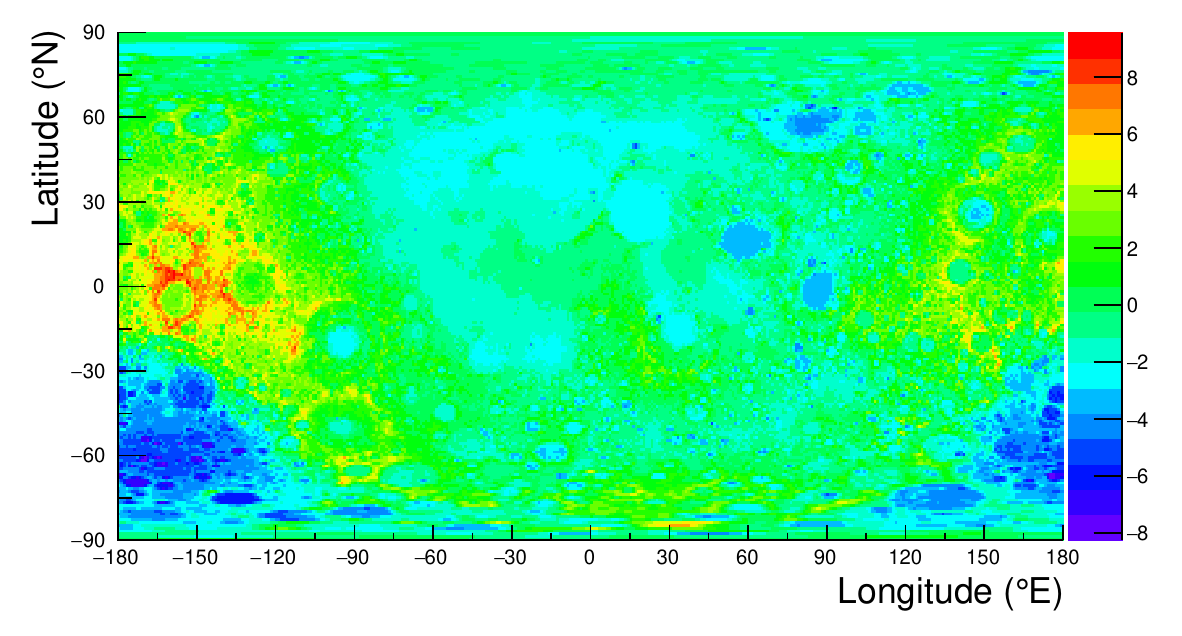}
\caption{Map of the lunar crust elevation (km).}
\label{fig:lunar_crust_elevation}
\end{figure}

The foundation of our physical model of the moon is a radial density profile \cite{briaud_etal2023}, which fixes the lunar crust according to the average of the four GRAIL-based models \cite{wieczorek_etal2013}, identifies a solid inner core, and confirms the existence of a low viscosity zone (LVZ) at the core-mantle boundary. Using this profile, we calculate the masses of the mantle and LVZ to be $(66.54\pm0.23)\times10^{21}$ kg and $(1.911\pm0.196)\times10^{21}$ kg, respectively. Figures~\ref{fig:lunar_crust_thickness} maps the thickness of the model-averaged lunar crust. The average thickness is $38.74$ km, the maximum thickness is $73.74$ km, and the minimum thickness is $1.05$ km.  Figure~\ref{fig:lunar_crust_density} maps the density of the model-averaged lunar crust. The density, which does not vary with depth, is the GRAIL-measured grain density reduced by a model-averaged porosity of 9.5\%. The average density is $2645$ kg/m$^3$ and the crust mass is $3.886\times10^{21}$ kg. Adding the masses of the LVZ, mantle, and crust, we get $(72.34\pm0.42)\times10^{21}$ kg for the mass of the bulk silicate moon (BSM). Table~\ref{tab:silicate_layer_masses} lists these masses. The absence of uncertainty on the crust mass is consistent with the radial density profile \cite{briaud_etal2023}. Figure~\ref{fig:lunar_crust_elevation} maps the elevation of the lunar crust relative to the constant $1737.151$ km lunar radius \cite{wieczorek_etal2013}. We use the elevation for calculating the antineutrino fluxes from the crust. The average elevation is effectively $0$ km, the maximum elevation is $9.56$ km, and the minimum elevation is $-8.26$ km.

\begin{table}
\caption{The masses of the silicate layers of the moon sum to give the BSM mass.}
\setlength{\tabcolsep}{2pt}
\begin{tabular}{l c c c c}
\hline\noalign{\smallskip}
                                           & LVZ                       & Mantle                & Crust       & BSM       \\
\hline\noalign{\smallskip}
Mass ($10^{21}$ kg)          & $1.91\pm.20$  & $66.54\pm.23$ & $3.89$  & $72.34\pm.42$ \\
\hline\noalign{\smallskip}
\end{tabular}
\label{tab:silicate_layer_masses}
\end{table}

\section{BSM Heat Flux}
We match the BSM abundances of thorium and uranium with those in the earth's primitive mantle \cite{mcdonough_sun_1995}. This establishes the heat-producing element (HPE) budget of the moon. Our approach, which traces the moon's origin to a fission fragment of the early earth \cite{daly_1946}, sets BSM uranium and thorium abundances equal to $20.3\pm4.1$ ppb and $79.5\pm11.9$ ppb, respectively. The giant-impact event significantly reduces the $240\pm48$ ppm of moderately volatile potassium in earth's primitive mantle. We assume $25\%$ or $60.0\pm12.0$ ppm of the ejected potassium finds its way into the moon. In our model, the lunar HPE masses are $(5.75\pm0.86)\times10^{15}$ kg Th, $(1.47\pm0.29)\times10^{15}$ kg U, and $(3.47\pm0.69)\times10^{18}$ kg K. The lunar $Th/U$ is $3.92\pm0.98$ and the lunar $K/U$ is $2956\pm836$. We only use the central values of the element ratios and list the uncertainties for completeness. Our estimate of the radiogenic heating in the BSM is $311\pm37$ GW, which leads to a surface-averaged heat flux of $8.19\pm0.97$ mW/m$^2$. These estimates assume that the uncertainties given for the earth's primitive mantle are uncorrelated.

\section{Crust Heat Flux}
\begin{figure}
\centering
\includegraphics[trim = 0mm 0mm 0mm 0mm, clip, scale=0.40]{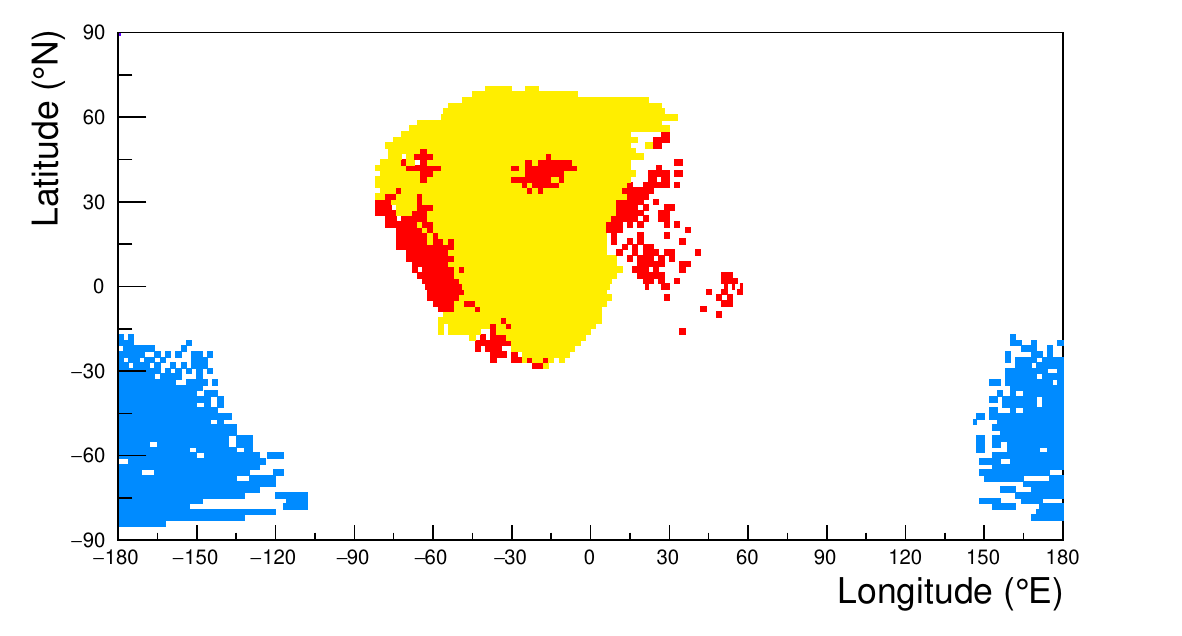}
\caption{Map of the lunar provinces: FHT in white, SPA in blue, PKT-Th in yellow, and PKT-FeO in red.}
\label{fig:lunar_terranes}
\end{figure}

We assign HPE abundances to stratified layers of the lunar crust across three provinces, following definitions for estimating the lunar bulk chemical composition \cite{taylor_wieczorek_2014}. The definitions of the three provinces, which are the Procellarum KREEP Terrane (PKT), the South Pole-Aitken (SPA) impact basin, and the Feldspathic Highlands Terrane (FHT), use surface thorium and ferrous oxide weight fractions measured by the Lunar Prospector gamma ray spectrometer \cite{prettyman_etal2006}. The PKT province has two regions; a contiguous region, exhibiting high surface thorium (Th $>4$ ppm) and a distributed region, exhibiting high surface ferrous oxide (FeO $>13$ wt\%) and elevated surface thorium (Th $>2.75$ ppm from the low altitude $2^\circ$ LP data). We label the contiguous region PKT-Th and the distributed region PKT-FeO but assign the same HPE contents to each. The SPA has elevated surface ferrous oxide (FeO $>6.5$ wt\%) and is bounded by latitude less than $-17^\circ$ N and by longitude greater than $146^\circ$ E and less than $-105^\circ$ E. The relatively thick and Th-poor FHT comprises the remainder of the lunar surface. Figure~\ref{fig:lunar_terranes} maps the three lunar provinces. An eastern archipelago of PKT-FeO, extending to the southeast, contains the site of the Apollo 17 heat flow measurement \cite{davies_etal1987}. The Apollo 15 heat flow measurement site lies inside the eastern PKT-Th \cite{davies_etal1987}. A third heat flow measurement site near the lunar south pole lies in the FHT \cite{wei_etal2023}. Table~\ref{tab:lunar_provinces} lists the characteristics of the provinces we use in our model of the lunar crust.

Starting from the surface and working downward, we recognize five stratified crustal layers: surface, top, middle, lower, and bottom. We characterize each layer by fractional thickness and thorium abundance. Uranium and potassium abundances simply scale from the lunar $Th/U$ and $K/U$. The top $1/30$ of the thickness of each of the gridded crust segments carries the surface abundance of thorium given by Lunar Prospector \cite{prettyman_etal2006}.  Lateral variations of uranium and potassium mirror those of thorium. An upper crust layer with $1/10$ of the thickness of each of the gridded crust segments has $0.5$ ppm thorium. A layer of ferroan anorthosite with $1/10$ of the thickness of each of the gridded crust segments has $0.05$ ppm thorium. The most massive crust layer, which is $11/15$ of the thickness of each of the gridded crust segments and labeled lower layer, has thorium abundances associated by province: $5$ ppm Th in PKT and SPA and $0.05$ ppm Th in FHT. The bottom $1/30$ of the thickness of each of the gridded crust segments has $5$ ppm thorium. Table~\ref{tab:layer_stuff} lists the fractional thickness and thorium content of the stratified crustal layers. 

\begin{figure}
\centering
\includegraphics[trim = 0mm 0mm 0mm 0mm, clip, scale=0.40]{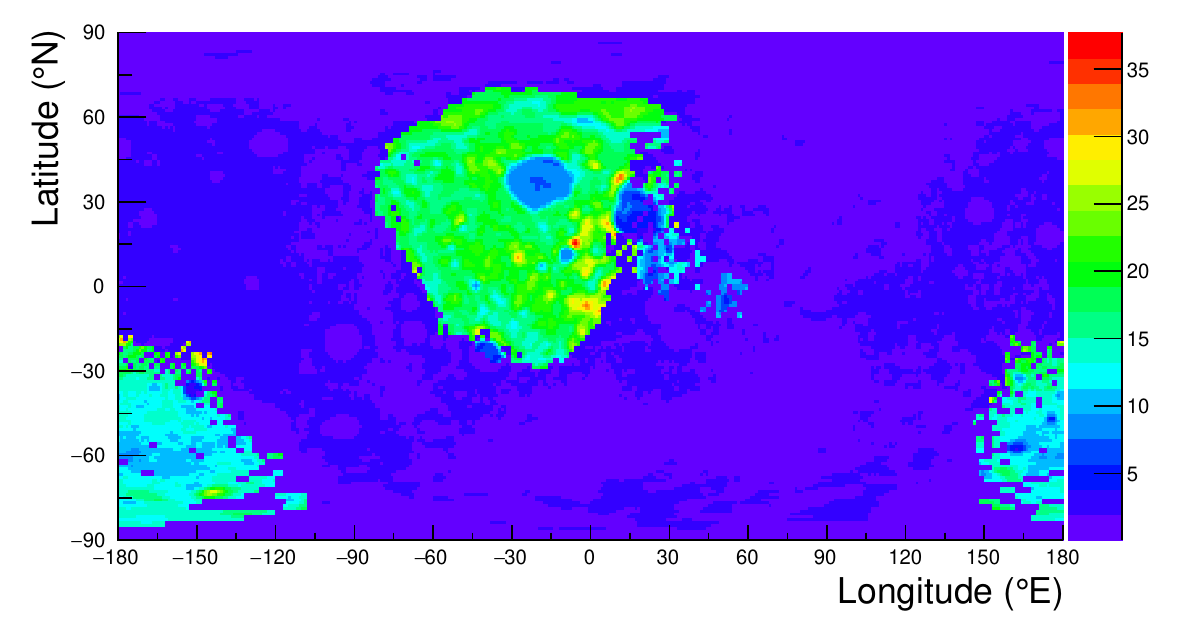}
\caption{Map of the lunar crust heat flux (mW/m$^2$).}
\label{fig:crust_heatflux}
\end{figure}

We arbitrarily assign $20\%$ uncertainty to HPE masses in the crust. Our crust model finds $189\pm35$ GW of radiogenic heating and a surface-averaged heat flux of $4.98\pm0.92$ mW/m$^2$. Assuming radially outward heat flow with no lateral dispersion, we find maximum and minimum crustal heat fluxes of $41.0\pm0.8$ mW/m$^2$ and $0.3\pm0.1$ mW/m$^2$, respectively. Figure~\ref{fig:crust_heatflux} maps the heat flux given by our crust model. Table~\ref{tab:lunar_provinces} summarizes our findings for the crust in each of the lunar provinces. 

\begin{table}
\caption{Characteristics of the stratified crust layers. Thorium abundances in the surface layer vary according to data from Lunar Prosector  \cite{prettyman_etal2006}. Thorium abundances in the lower layer, which depend on the lunar province, are $5$ ppm in PKT, $5$ ppm in SPA, $0.05$ ppm in FHT.}
\setlength{\tabcolsep}{2pt}
\begin{tabular}{l c c c c c}
\hline\noalign{\smallskip}
                                           &  Surface   & Upper      & Middle       & Lower              & Bottom   \\
\hline\noalign{\smallskip}
Fractional thickness           & $1/30$     & $1/10$     & $1/10$      & $11/15$            & $1/30$ \\ 
Th (ppm)                            & varies      & $0.5$       & $0.05$      &  $5$ or $.05$   &  $5$      \\
\hline\noalign{\smallskip}
\end{tabular}
\label{tab:layer_stuff} 
\end{table}

\begin{table}
\caption{Characteristics of the lunar crust in each of the lunar provinces.}
\setlength{\tabcolsep}{2pt}
\begin{tabular}{l c c c c}
\hline\noalign{\smallskip}
                                               &  PKT-Th   & PKT-FeO  & SPA         &  FHT   \\
\hline\noalign{\smallskip}
Area ($10^6$ km$^2$)           & $4.67$     & $1.25$     & $1.96$      & $30.0$  \\ 
Avg. thickness (km)                & $31.4$     & $26.9$     & $23.9$      & $41.3$  \\ 
Avg. surface Th (ppm)            & $6.61$     & $3.87$     & $2.88$      & $1.66$  \\
Heating (GW)                         & $87.0$     & $19.5$     & $27.1$      & $55.1$  \\
Avg. heat flux (mW/m$^2$)    & $18.6$     & $15.5$     & $13.9$      & $1.83$  \\
Max. heat flux (mW/m$^2$)    & $37.7$     & $32.8$     & $33.1$     & $3.20$  \\
Min. heat flux (mW/m$^2$)    & $6.13$     & $4.44$      & $2.56$     & $0.05$  \\
\hline\noalign{\smallskip}
\end{tabular}
\label{tab:lunar_provinces}
\end{table}

\section{Mantle and LVZ Heat Fluxes}
Taking the HPE masses given by our crust model, and by mass balance with the BSM amounts, we find the HPE masses leftover for the mantle plus LVZ. Table~\ref{tab:HPE_masses} lists the HPE masses, the radiogenic heating, and the surface-averaged heat flux for the BSM, the crust, and the mantle plus LVZ. The radiogenic heating and the surface-averaged heat flux from the mantle plus LVZ are $127\pm14$ GW and $3.34\pm0.37$ mW/m$^2$, respectively.

The existence of a LVZ, as identified in the radial density profile \cite{briaud_etal2023}, supports the scenario of deep lunar mantle overturn. In this process dense, HPE-rich cumulate at the surface sinks down to the top of the mantle \cite{liang_etal2024}. We approximate this by increasing the HPE abundances in the LVZ by $10$ times over those in the mantle. This partitioning puts $27.3\pm0.26$ ppb Th and $6.98\pm0.66$ ppb U in the mantle. The resulting HPE masses are $(1.82\pm0.17)\times10^{15}$ kg Th, $(4.65\pm0.44)\times10^{14}$ kg U, and $(1.37\pm0.13)\times10^{18}$ kg K in the mantle, and $(5.23\pm0.49)\times10^{14}$ kg Th, $(1.33\pm0.13)\times10^{14}$ kg U, and $(3.94\pm0.37)\times10^{17}$ kg K in the LVZ. As long as the HPE abundances are uniformly distributed in these deep silicate shells, partitioning between shells has no effect on our estimates of the surface heat flux.

\begin{table}
\caption{The HPE masses, radiogenic heating, and surface-averaged heat fluxes for the BSM, the crust, and the mantle plus LVZ from our lunar model, using mass balance constraints.}
\setlength{\tabcolsep}{2pt}
\begin{tabular}{l c c c}
\hline\noalign{\smallskip}
                                               &  BSM                  & Crust                    &  Mantle+LVZ       \\
\hline\noalign{\smallskip}
Th ($10^{15}$ kg)                  & $5.75\pm0.86$    & $3.41\pm.64$      & $2.34\pm0.22$   \\
U ($10^{15}$ kg)                    & $1.47\pm0.29$    & $0.87\pm.16$      & $0.60\pm0.13$   \\
K ($10^{18}$ kg)                    & $4.34\pm0.87$    & $2.57\pm.49$      & $1.77\pm0.38$   \\
Heating (GW)                        & $311\pm37$        & $184\pm23$         & $127\pm14$       \\
Heat flux (mW/m$^2$)          & $8.19\pm0.97$    & $4.86\pm0.62$     & $3.34\pm0.37$   \\
\hline\noalign{\smallskip}
\end{tabular}
\label{tab:HPE_masses}
\end{table}

We sum the variable heat flux from the crust with the constant heat flux from the mantle plus LVZ to find the total surface heat flux. We compare our results with measurements at three sites: Apollo 15, Apollo 17, and Haworth. As mentioned above, the Haworth and Apollo 15 heat flow measurement sites lie inside the FHT and the PKT provinces, respectively. However, the Apollo 17 heat flow measurement site lies within the PKT archipelago extending into the FHT province. We find the lowest heat fluxes emerging from the FHT province and the highest heat fluxes emerging from the PKT province. Areas with FHT and PKT crust segments mixed together, like the archipelago, are prone to lateral heat flow. Heat from the warmer PKT crust segments could flow laterally to the neighboring, cooler FHT crust segments. We take the liberty of averaging the heat fluxes from the FHT and PKT provinces for the crust segment containing the Apollo 17 site by assigning $2.5$ ppm Th to the lower crust. Table~\ref{tab:heatflow_sites} compares the heat fluxes from our model with measured values at several sites. Our model, which constrains the HPE content of the BSM with the earth's primitive mantle and recognizes three provinces (PKT, SPA, FHT) with distinct distributions of HPEs, finds heat fluxes in agreement with the measurements at all sites.

\begin{table}
\caption{Our modeled heat fluxes compared with measurements at the Apollo 15 and Apollo 17 sites \cite{davies_etal1987} \cite{langseth_etal1976} and at Haworth \cite{wei_etal2023}, a permanently shadowed crater near the lunar south pole. Our model heat flux at the Apollo 17 site uses thorium abundance in the lower crust layer set to $2.5$ ppm.}
\setlength{\tabcolsep}{2pt}
\begin{tabular}{l c c c}
\hline\noalign{\smallskip}
                                             &  Apollo 15             & Apollo 17            &  Haworth       \\
\hline\noalign{\smallskip}
Latitude ($^\circ$N)               & $26.1$                & $20.2$                &  $-87.5$           \\ 
Longitude ($^\circ$E)            & $3.6$                  & $30.8$                &   $-10$             \\
Crust thickness (km)             & $34.3$                & $39.2$                &   $31.8$          \\
Our model (mW/m$^2$)        & $23.6\pm4.4$    & $15.4\pm2.8$      &   $4.6\pm0.6$  \\
Published (mW/m$^2$)         & $21\pm3$          & $15\pm2$           &   $4.9\pm0.2$  \\ 
\hline\noalign{\smallskip}
\end{tabular}
\label{tab:heatflow_sites}
\end{table}

\section{Antineutrino fluxes}

We estimate the antineutrino fluxes from the lunar crust by numerical integration. Each of the $86400$ $1^\circ$ by $1^\circ$ gridded crustal source segments is divided vertically into $30$ slices of equal thickness. The field point for the estimated flux is always at the surface and in the center of a crust segment. If the distance between the field point and the middle of the crustal source slice is less than $400$ km, then the slice is divided into $36$ source pieces by making six evenly-spaced cuts in both latitude and longitude.

Our numerical integration finds average crustal antineutrino fluxes of $4.00\times10^5$ cm$^{-2}$s$^{-1}$ from $^{232}$Th, $4.65\times10^5$ cm$^{-2}$s$^{-1}$ from $^{238}$U, $1.45\times10^4$ cm$^{-2}$s$^{-1}$ from $^{235}$U, $4.59\times10^5$ cm$^{-2}$s$^{-1}$ from $^{40}$K$_\beta$ the beta decay branch of potassium, and $6.61\times10^3$ cm$^{-2}$s$^{-1}$ from $^{40}$K$_\mathrm{ec}$ the electron capture decay branch of potassium. Collectively, these crustal contributions sum to an average antineutrino flux of $1.34\times10^6$ cm$^{-2}$s$^{-1}$. The uncertainty of our estimates for the crustal antineutrino fluxes is $20\%$, as arbitrarily assigned to the crustal HPE abundances.

\begin{figure}
\centering
\includegraphics[trim = 0mm 0mm 0mm 0mm, clip, scale=0.40]{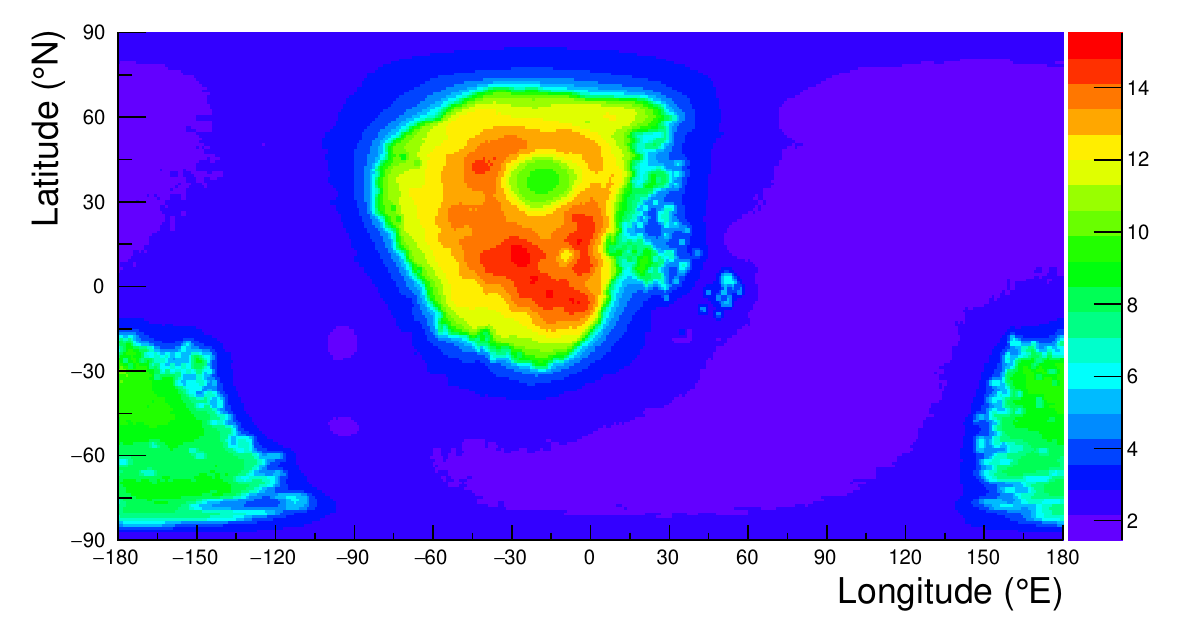}
\caption{Map of the lunar crust antineutrino inverse beta decay rate (NIU).}
\label{fig:lunar_crust_NIU}
\end{figure}

Figure~\ref{fig:lunar_crust_NIU} maps the rate of inverse beta decay (IBD) interactions on free protons due to antineutrinos from uranium and thorium in the lunar crust. The average crustal IBD rate is $4.30\pm0.86$ interactions per $10^{32}$ targets per year, which is hereafter called a neutrino interaction unit (NIU). All IBD rates assume an average neutrino oscillation survival probability of $0.5529$. The maximum and minimum crustal IBD rates are $15.5\pm3.1$ NIU and $1.5\pm0.3$ NIU, respectively. We note that the ratio of interactions from uranium and thorium over the lunar surface in our model is constant. This is potentially at odds with the data from Kaguya \cite{yamashita_etal2010}.

The geophysical responses at the lunar surface for the spherically symmetric mantle and LVZ are $(2.501\pm0.001)\times10^8$ g/cm$^2$ and $(5.173\pm0.068)\times10^6$ g/cm$^2$, respectively. The uncertainties on the geophysical responses arise only from the densities of the shells in the radial density profile \cite{briaud_etal2023}. Calculation of the geophysical response at a field point outside of a spherically symmetric shell is possible in closed form \cite{kgs84}. Our estimate of the surface antineutrino flux from the mantle plus LVZ is $(4.89\pm0.46)\times10^5$ cm$^{-2}$s$^{-1}$. The resulting IBD rate from the mantle plus LVZ is $1.48\pm0.14$ NIU. For comparison, in the special case of equal HPE abundances distributed uniformly throughout both the mantle and LVZ the IBD rate is $1.57\pm0.15$ NIU.

\section{Discussion}
Our lunar model is available online at https://reactors.geoneutrinos.org. The web application has controls for adjusting the HPE abundances in the mantle plus LVZ. There are options for changing the antineutrino interaction type, either inverse beta decay or electron elastic scattering. Moreover, the rate of CEvNS interactions are available for various selectable elements. Plots, including those in this paper as well as the interaction rate energy spectrum, are interactive. There are download buttons for the energy spectra data as well as the plots. Finally, there is the option for adding the antineutrino interactions from the core of a nuclear reactor with a designated location, thermal power, and fuel type.

NASA's Artemis program \cite{creech_etal2022} aims to establish a lunar base early next decade. The current plan deploys the base in the south polar region, near permanently shadowed regions, which are thought to contain water ice. For energy, the plan calls for small nuclear fission reactors, producing $120$ kW$_{\mathrm{th}}$ fueled by low enriched uranium (LEU). The mass limit target for the reactor is less than $6$ metric tons, which we assume to be the payload of the delivery rocket. At a distance of $10$ m from a $120$ kW$_{\mathrm{th}}$ core burning LEU fuel the IBD rate from reactor antineutrinos is roughly $1$ interaction per metric ton of water per day ($359\pm10$ per year). A detector with a $6$ metric ton water target and $50$\% efficiency would record $\sim3$ events per day. With sufficient shielding and background rejection, this rate is sufficient for monitoring the reactor operational state. In comparison, the maximum IBD rate from primordial radioactivity within the moon, which is $\simeq17$ NIU equals the rate from a $120$ kW$_{\mathrm{th}}$ core at a standoff distance of approximately $1.7$ km. For the minimum IBD rate from primordial radioactivity within the moon, which is $\simeq3$ NIU, the equal rate standoff distance increases to approximately $4.2$ km.

\section{Conclusion}
We present estimates for the fluxes of heat and antineutrinos due to primordial radioactivity within the moon. We find a heat flux of $8.19\pm0.97$ mW/m$^2$, an antineutrino flux of $(1.83\pm0.32)\times10^6$ cm$^{-2}$s$^{-1}$, and an IBD interaction rate of $5.8\pm1.0$ NIU, all averaged over the lunar surface. Our estimates of the heat flux over the lunar surface could potentially help in selecting the location of future lunar heat flow measurements or sample collection sites. Our estimates of the antineutrino fluxes allow calculation of interaction rates from various detector targets, including free protons, atomic electrons, and selected atomic nuclei. Calculated interaction rates could potentially help design antineutrino detection projects with applications for testing models of the moon's origin and thermal evolution, as well as for extraterrestrial nuclear monitoring and security.

\newpage
\section*{Acknowledgments}
We are thankful for discussions with G. Jeffrey Taylor, whose generous guidance greatly improved our lunar model. This work was supported in part by Lawrence Livermore National Security, LLC.

\newpage
\bibliographystyle{apsrev4-1}
\nocite{apsrev41control}
\bibliography{Lunar_bib,revtex-custom}

\end{document}